\documentclass[10.5pt]{article}
\pdfoutput=1
\usepackage{multicol}
\usepackage[T1]{fontenc}
\usepackage[utf8]{inputenc}
\usepackage[affil-it]{authblk}
\usepackage[fleqn]{amsmath}
\usepackage{enumitem}
\setlist[enumerate]{noitemsep}
\setlist[enumerate]{topsep=4pt}
\usepackage{amssymb}
\usepackage{graphicx}
\usepackage{caption}
\usepackage{subcaption}
\usepackage{geometry}
\usepackage{titlesec}
\titlespacing*{\section}
{0pt}{5.5ex plus 1ex minus .2ex}{4.3ex plus .2ex}
\titlespacing*{\subsection}
{0pt}{3ex plus 1ex minus .2ex}{2ex plus .2ex}
\newcounter{tempcolnum}
\makeatletter
\newcommand{\multicolinterrupt}[1]{
\setcounter{tempcolnum}{\col@number}
\end{multicols}
#1%
\begin{multicols}{\value{tempcolnum}}
}
\makeatother

\title{Addressing current challenges in cancer immunotherapy with mathematical and computational modeling}
\author[1]{Anna Konstorum}
\author[2]{Anthony T. Vella}
\author[2]{Adam J. Adler}
\author[3]{Reinhard Laubenbacher\thanks{Corresponding author: laubenbacher@uchc.edu}}
\affil[1]{Center for Quantitative Medicine, UConn Health, Farmington, CT}
\affil[2]{Department of Immunology, UConn Health, Farmington, CT }
\affil[1,3]{Jackson Laboratory for Genomic Medicine, Farmington, CT}
\date{}

\begin{document}
\maketitle
\begin{abstract}
\noindent The goal of cancer immunotherapy is to boost a patient's immune response to a tumor.  Yet, the design of an effective immunotherapy is complicated by various factors, including a potentially immunosuppressive tumor microenvironment, immune-modulating effects of conventional treatments, and therapy-related toxicities.  These complexities can be incorporated into mathematical and computational models of cancer immunotherapy that can then be used to aid in rational therapy design.  In this review, we survey modeling approaches under the umbrella of the major challenges facing immunotherapy development, which encompass tumor classification, optimal treatment scheduling, and combination therapy design.  Although overlapping, each challenge has presented unique opportunities for modelers to make contributions using analytical and numerical analysis of model outcomes, as well as optimization algorithms.  We discuss several examples of models that have grown in complexity as more biological information has become available, showcasing how model development is a dynamic process interlinked with the rapid advances in tumor-immune biology.  We conclude the review with recommendations for modelers both with respect to methodology and biological direction that might help keep modelers at the forefront of cancer immunotherapy development. \\
\end{abstract}
\begin{multicols}{2}
\section{Introduction}
The involvement of the immune system in all stages of the tumor lifecycle, including prevention, maintenance, and response to therapy is now recognized as central to understanding cancer development from a systemic point of view.  Therefore, it is not surprising that one of the most rapidly developing and exciting fields in cancer treatment is that of cancer immunotherapy, i.e. therapy that boosts the function of the patient's own immune system in targeting the cancer.  The development of a knowledge-base of tumor-immune interactions and basic and clinical work on cancer immunotherapy has been paralleled by the development of mathematical and computational models that utilize this knowledge-base to design \emph{in silico} model systems upon which immune-based and other treatments can be modeled.  In the best case scenario, these models can serve to guide clinicians and developers of clinical trials towards optimizing mono- and combination therapies and basic scientists in understanding the underlying mechanisms of the effectiveness (or, ineffectiveness) of therapy combinations.  Therefore, in a field as rapidly developing and clinically important as cancer immunotherapy, mathematical and computational modeling can play a central role in helping to guide the direction the field takes. 

\indent In this review, we survey the mathematical modeling work in cancer immunotherapy organized by the 'major challenges' that the immunotherapy community is currently grappling with.  We will also outline strategies that modelers can use to further enhance their contribution to addressing these challenges. 

This review is organized as follows.  In Section \ref{sec:challenges}, we give a brief overview of tumor immunology and cancer immunotherapy.  We also outline `major challenges' that have been posed in the immunotherapy community.  In Sections \ref{sec:class}-\ref{sec:comb}, we survey how each challenge has been addressed by the modeling community; in Section \ref{sec:recs}, we provide recommendations as to what techniques the community can employ to further progress on each challenge and to address other rising challenges, and in Section \ref{sec:conc}, we summarize our findings.\vspace{-0.2in}   
\section{The major challenges for cancer immunotherapy}
\label{sec:challenges}

A tumor can, in principle, be recognized and controlled by the patient's immune system via a coordinated process that has recently been summarized as the `cancer-immunity' cycle \cite{Chen:2013kx}.  Dying tumor cells release proteins that contain unique tumor-specific antigens that are either mutated or differentially modified post-translationally relative to normal cells \cite{Finn:2008vn}.  Tumors also release inflammatory signals in the form of cytokines and other factors (such as heat-shock proteins) that lead to a local innate inflammatory response.  Dendritic cells (DCs), which form a part of this response, can uptake these antigens and, if properly activated by other factors in the tumor microenvironment, differentiate to present these antigens to lymphoid cells via their MHC class I and II molecules.  Activation of cytotoxic CD8$^+$ T cells results in their proliferation and trafficking to the tumor site in order to kill the tumor cells, leading to the release of more tumor-associated antigens and thus repeating the cycle.  This process by which the immune system keeps a tumor in check is defined as cancer immunosurveillance.  Via a counter-process termed immunoediting, a tumor can evolve and strengthen various mechanisms to escape immunosurveillance \cite{Dunn:2002ys, Schreiber:2011uq, Mittal:2014fk}.  

The goal of cancer immunotherapy is to boost the response of the immune system to the tumor by intervening at one or several points of the cancer-immunity cycle.  The antigen-based activation of dendritic cells can be achieved by administration of a therapeutic vaccine harboring one or multiple tumor-associated antigens, with or without additional factors that prime dendritic cells for activation.  Another approach has been to extract peripheral blood monocytes and stimulate them to become DCs via presentation of antigens plus costimulatory molecules \emph{ex vivo} before reintroducing them into the patient.  An FDA-approved example of the latter is the drug Provenge (sipuleucel-T; Valeant Pharmaceuticals), which improves median survival time in advanced prostate cancer by $\sim4\%$ \cite{Kantoff:2010zr}.  Therapies that directly boost  anti-tumor T cell activity include adoptive T cell therapy and inhibition of T cell checkpoint molecules.  In adoptive cell therapy (ACT), T cells are collected from a patient, expanded \emph{ex vivo}, and reintroduced into the patient.  This method has been highly successful in melanoma, and is being explored for other cancers \cite{Rosenberg:2008fk}.  Checkpoint therapy involves antagonizing  T cell inhibitory receptors (such as CTLA-4 or PD-1). Reciprocally, there are a number of activating costimulatory receptors on T cells (such as OX40, 4-1BB, GITR, and CD27) that can be engaged to boost T cell clonal expansion and acquisition of tumoricidal effector functions \cite{Mellman:2011kx}.  Major successes using this method include the FDA-approved Yervoy (ipilimumab; Bristo-Myers Squibb), a monoclonal antibody (mAb) to CTLA-4 that is used for patients with melanoma \cite{Weber:2007uq, Hodi:2010kx}.  In 2016, the FDA approved Tecentriq (atezolizumab; Genentech), an inhibitor of the ligand for PD-1 (PD-L1) that is expressed on tumor cells and that would otherwise engage PD-1 on tumor-infiltrating T cells, for treatment of locally advanced or metastatic bladder cancer \cite{Tarapchak:2016}.  For a more extensive discussion of cancer immunotherapies and drugs in clinical development, see \cite{Mellman:2011kx, Chen:2013kx, Jeanbart:2015ys}.

There are particular challenges that present themselves in almost all applications of cancer immunotherapy, and have been addressed by the mathematical community.  These major challenges include: 
\begin{enumerate}
\item Tumor classification for treatment and prediction of response.
\item Optimal scheduling and dosage of treatment.
\item Design and identification of combination treatment regimes.
\end{enumerate}
In this review, we use these `major challenges' as a means by which to present how mathematical modeling can help to address each challenge and thereby help to progress the field of immunotherapy research and application.   A summary of the tumor-immune and immunotherapy interactions consistently addressed by the models across these challenges is depicted in Figure \ref{fig:summary}.  \\

\multicolinterrupt{
\begin{center}
\includegraphics[width=.8\linewidth]{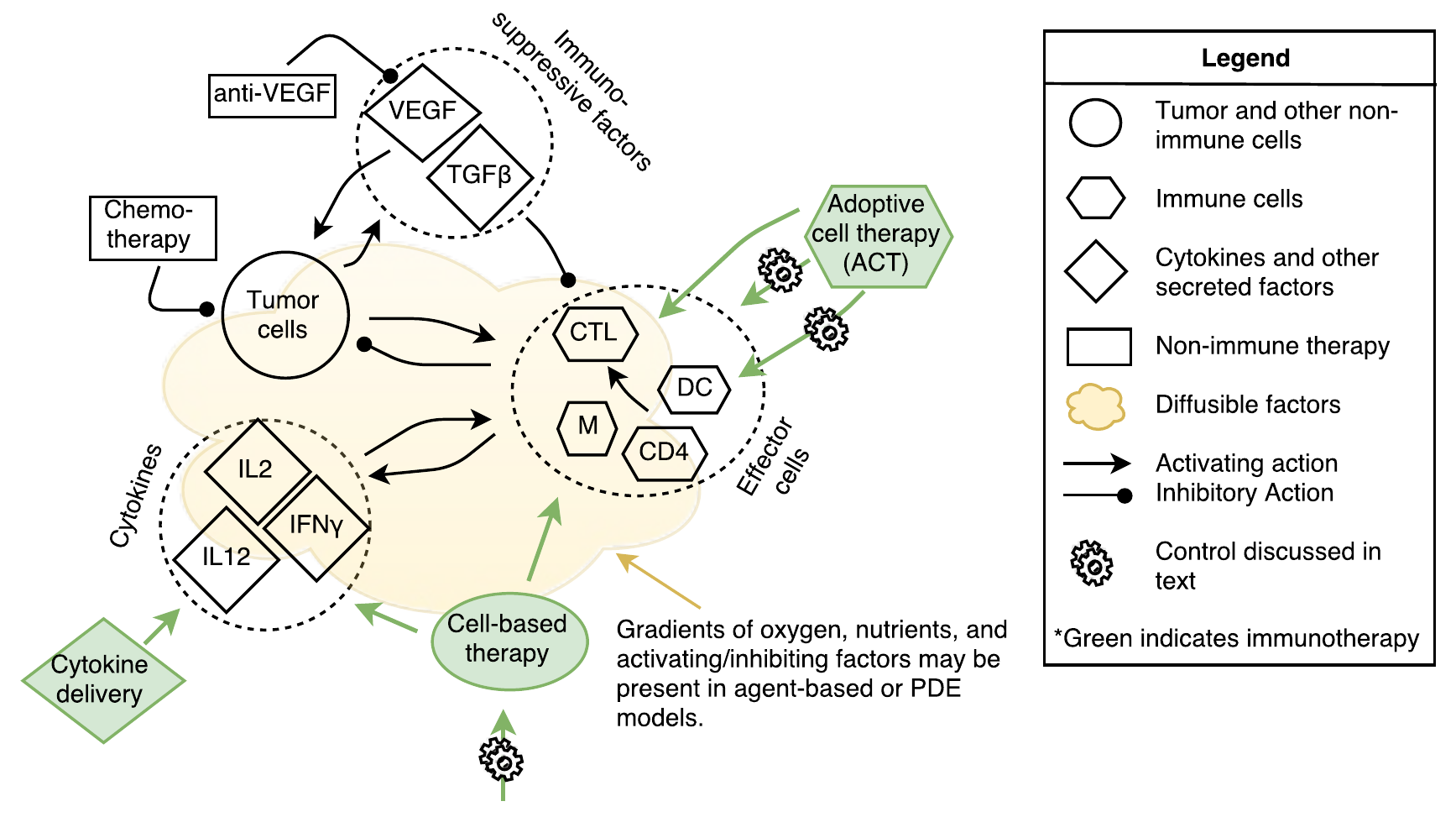}
\captionof{figure}{ \footnotesize Summary of modeling efforts in immunotherapy.  Generally, models (both mathematical and computational) are at the cell-population level and consist of tumor and immune cells (immune cells may be represented cumulatively as effector cells or more specifically by cell-types), cytokines, and other immune-activating or inhibiting factors such as VEGF or TGF$\beta$, with therapies (in green) targeting any of these factors.  Models can assume spatial homogeneity (ordinary differential equation, ODE models) or be spatially heterogeneous (agent-based and partial differential equation (PDE) models).  For addressing tumor classification for treatment (Challenge 1, Section \ref{sec:class}), analytical, numerical, and/or parameter sensitivity analysis is performed on the models to determine which system parameters contribute most strongly to prediction of treatment response.  Control theory and evolutionary algorithms (depicted by cogwheels) are applied to modeled treatments  to optimize scheduling and dosage (Challenge 2, Section \ref{sec:optimal}), and all these methods can be applied to multiple treatments, including non-immune therapy such as chemo- and molecular therapy, to model combination treatment regimes (Challenge 3, Section \ref{sec:comb}).  Abbreviations: CTL, cytotoxic T cells; DC, dendritic cells; CD4, CD4$^+$ T cells, M, macrophages. }
\label{fig:summary}
\end{center}
}
\section{Challenge: Tumor classification for treatment and prediction of response}
\label{sec:class}
Tumor classification, usually based on histopathological grading, does not serve as a strong predictive tool for post-treatment outcome.  More sophisticated methods of tumor classification, such as immune profiling \cite{Galon:2012vn}, inclusion of markers of the tumor microenvironment \cite{Overacre:2015}, and incorporation of high-throughput data \cite{Srivastava:2013ve}, can improve the predictive strength of the classifications.  Nevertheless, due to both the increasing availability of patient data and treatment options, it can become difficult to predict how a patient with a specific set of tumor characteristics will respond to a given treatment.  Modeling efforts in this regard have been used to identify which patient parameters may be most important to predicting therapy outcomes.  Since mathematical models, unlike statistical models, are typically mechanistic, they can be used to predict the effect of therapy or therapy combinations that have not yet been tried in the clinic (i.e. for which patient data is as yet unavailable). 
We begin with the Panetta-Kirschner (PK) model  \cite{Kirschner:1998fk}, one of the first to layer immunotherapy into a tumor-immune model.  The PK model consists of system of three ordinary differential equations (ODEs) that model the dynamics of effector ($E$) and tumor ($T$) cells, and the cytokine IL-2 ($I_L$).  For the sake of expediency and since one of the parameters is especially important in the analysis of model behavior, we only show the equation for E, 
\begin{equation}
\frac{dE}{dt} = cT - \mu_2 E + \frac{p_1 E I_L}{g_1 + I_L} + s_1. \label{eqn:PK_E}
\end{equation}

The antigenecity of the tumor is modeled by the parameter $c$, $\mu_2$ represents the death rate of $E$, and the proliferative effect of IL-2 on $E$ is in Michaelis-Menton form to model saturation of response.  Therapies are considered by terms $s_1$ and $s_2$ added to the right-hand side of $d E / d t$ and $d I_L / d t$, respectively.  Michaelis-Menton terms are also used to model the negative effect of $E$-induced tumor destruction on $T$, and the positive effect of tumor-resident $E$ on $I_L$ production.  Tumor growth is modeled using the logistic growth function.  The authors show that with no therapy, if $c$ is below a critical $c_0$, the only stable steady state is a large tumor.  As $c$ increases, the tumor size oscillates between large and small, with the time it spends in its 'large' state (and the magnitude of the 'large' state) decreasing with increasing $c$.  Adding therapy $s_1$ creates a tumor-free equilibrium that is stable if $s_1 > s^1_{\text{crit}}$ where $ s^1_{\text{crit}}$ depends on several parameters of the model.  A bifurcation diagram of $s^1_{\text{crit}}$ vs. $c$ shows that there exist regions where the tumor will either die or survive depending on $c$, hence providing a mathematical basis for tumor classification for treatment.  The authors show similar results for $s_2>0$ and both $s_1$ and $s_2>0$.  

Models with a relatively small number of equations can be analyzed in this way and have consistently shown that knowledge of system parameters can help lead to tumor classification for treatment via bifurcation and stability analysis (e.g. \cite{Nani:2000fk, Byrne:2004, B_M:2007fk, Konstorum:2016cr}).  Identification of parameters where thresholds exist with respect to system response to therapy can aid in the identification of effective therapies.  For example, for the system modeled in  \cite{Kirschner:1998fk}, a therapy that increases the antigenecity, $c$, of the tumor can push the tumor past the critical point necessary for response to other therapies and/or reduction in size to a small, stable steady state.  An excellent review of simpler tumor-immune models and their associated analyses, which may or may not incorporate immunotherapy, can be found in \cite{Eftimie:2011kx}.  A meta-modeling approach was taken by d'Onofrio et al.\ \cite{dOnofrio:2008} to synthesize the analytical results of such models.  These models can serve as baseline models that can be extended to incorporate immunotherapy.

Numerical analysis of models has also focused on the concept of thresholds for predicting patient response.  For example, Kronik et al.\ \cite{Kronik:2012ve} set out to answer the question whether `mathematical modeling would help to define the prerequisites of an effective immunotherapy approach'.  They extended a previously published model of T cell transfer immunotherapy for glioblastoma \cite{Kronik:2008uq} to a model of \emph{ex vivo} expanded tumor-specific T cell transfer for melanoma and used clinical data to retroactively validate it.  The model consists of a system of ODEs with 5 equations representing tumor and immune cells, and critical signaling molecules produced by the two cell populations (TGF$\beta$, IFN$\gamma$, and MHC Class I molecules).  The authors varied initial tumor size and growth rate to imitate a virtual population of patients with heterogeneous tumor profiles.  Four different T-cell therapy regimens were then simulated over this population that corresponded to four different clinical trials.  For one such trial, the model was able to offer an explanation for the lack of a dose-response relationship with therapy: the model showed that for patients with large enough tumors, the therapy would have no effect, whereas in smaller tumors a dose-response relationship could be identified, again pointing to a threshold for therapy effectiveness, this time via simulation in lieu of a bifurcation analysis.  Indeed, the authors suggested that volumetric tumor analysis is a better predictor for clinical effectiveness of T cell therapy than the traditional staging method.  The authors also called for large doses of T cell therapy due to presence of thresholds for efficacy of T cell killing with respect to T cell concentrations, although this recommendation should be considered in light of potential toxicities.  More broadly, parameter sensitivity analysis has played a critical role in the numerical analysis of models of immunotherapy (e.g., \cite{dePillis:2005zr, B_M:2008uq, Joshi:2009ly, Banerjee:2015vn}) by helping to elucidate which tumor characteristics are most predictive of therapy success.

To investigate how spatial organization of different cell types can impact tumor-immune evolution and response to therapy, Wells et al.\ \cite{Wells:2015} developed a hybrid discrete-continuous (HDC) agent-based model.  These models treat cells as agents that sit on a lattice and can interact with and respond to other cells in a stochastic manner through growth factors and cytokines that can diffuse throughout the lattice.  The cell types in this model were naive, M1 ('tumor-suppressive'), and M2 ('tumor-promoting') macrophages, and live and dead tumor cells.  In addition to secreted factors, oxygen diffusion was also implemented in the model, thereby creating oxygen-rich and hypoxic regions in the tumor and its microenvironment.  The authors used a multiparametric sensitivity analysis (MPSA)  to observe that the ratio of M2 to other cell types (termed the Macrophage Polarization Index, MPI) early in the simulation is predictive of tumor survival.  Another key predictive outcome of the model was that increased tumor heterogeneity is  associated with tumor survival.  Tumor heterogeneity was associated with locally elevated stimulated macrophages, leading to local peaks of M2 differentiation which increased overall MPI.  It followed that a decrease in secretion of differentiation factors for M2 cells abrogated the association between tumor heterogeneity and survival in the simulation.  Therefore, the model provided a novel hypothesis for tumor-induced immunosuppression via increases in tumor-heterogeneity, but also a potential new prognostic tool for tumors where biopsies and cell-specific stains are available.  This observation was further used by the authors to hypothesize and test \emph{in silico} that  introduction of genetically engineered macrophages that could block the M2 transition would be effective in killing the tumor by blocking the feedback loop that generates localized sites of highly concentrated immunosuppressive M2 cells, thereby constructing a novel treatment using their observations of the most critical contributors to tumor growth.

Lattice- and/or agent-based models (ABMs) can provide additional information regarding spatially explicit parameters influencing tumor-immune interactions and response to immunotherapies.  In addition to \cite{Wells:2015}, this approach has been used by Papalardo et al.\ \cite{Pappalardo:2005kx, Palladini:2010ys} to develop an agent-based simulator of the Triplex vaccine, SimTriplex, which is effective in preventing mammary carcinoma in HER-2/neu transgenic mice, and Dr\'{e}au et al.\ \cite{Dreau:2009} to examine the interactions between a vascularized tumor and the immune response.  While partial differential equation (PDE) models have been much more sparsely employed to model immunotherapy, they can be especially useful when modeling a large number of interacting cells in a tissue, in which case building a corresponding ABM would be too computationally costly. For example, Eikenberry et al.\ \cite{Eikenberry:2009qf} developed a PDE of melanoma with immune infiltrate, and showed that surgical removal of primary tumors with high levels of immune infiltrate could promote growth of satellite metastases, as was observed clinically, thereby providing a model-based hypothesis for tumor classification with respect to responsiveness to therapy (in this case, surgery).
\section{Challenge: Optimal scheduling and dosage of treatment }
\label{sec:optimal}
For any given treatment, an exhaustive experimental search for optimal dosage and/or scheduling is unrealistic.  There are several techniques which developers of mathematical models for immunotherapy have used to identify optimal treatment schedules \emph{in silico}, including optimal control and genetic algorithms.
\subsection{Optimal control theory}
Optimal control theory has played a prominent role for using mathematical models of cancer immunotherapy for design of optimal therapy regimes.  We can consider a system of ordinary differential equations, 
\begin{align}
\begin{cases}
&\dot{\* x }  = \*f(\*x(t),\*u(t)) \hspace{0.1in}(t>0) \\
&\*x(0)  = x^0 
\end{cases}
\end{align}
where $\*x \in \mathbb{R}^n$ and $\*u \in \*U \subset \mathbb{R}^m$ is a control parameter (e.g. $x(t)$ can represent concentrations of tumor cells, immune cells and cytokine and $u(t)$ can represent an immunotherapy treatment) and an initial condition $x^0$.  The optimal $\*u$, $\*u^*$, is chosen from a space $\*U$ of admissible controls based on constraints on $\*x$ and $\*u$ via calculation of an extremal of an objective functional
\begin{equation}
J(\*u) = \int_0^T r(\*x(t), \*u(t))dt + g(\*x(T)) \label{eqn:Bolza},
\end{equation}
where $r$ is termed the \emph{running payoff} and $g$ the \emph{terminal payoff} \cite{Evans:Cont}.  This formulation is termed the \emph{Bolza} form.  The optimal $\*u^*$ found using this method then represents the best possible therapy given conditions on the control parameters (such as minimization of tumor cells, minimization of therapy dosage, maximization of immune response, etc.).

Optimal control theory has been applied extensively to the Panetta-Kirschner (PK) model \cite{Kirschner:1998fk}, described in Section \ref{sec:class}.  Burden et al.\ \cite{Burden:2004} made ACI therapy in the PK model into a control parameter, $u$, and built an objective functional
\begin{equation}
J_B(u) = \int_0^T[x(t) - y(t) + z(t) - \frac{1}{2} B\cdot u(t)^2] dt,
\end{equation}
where $x(t)$ are activated effector cells, $y(t)$ are the tumor cells, $z(t)$ the local concentration of IL-2, and $B$ the strength of patient tolerance to treatment (i.e., $B$ is inversely correlated to side effects of treatment).  For $t\in [0,T]$, The class of admissible controls was set to 
\[ U = \{ u(t) \mbox{ piecewise continuous } | 0 \le u(t) \le 1\} \]
and $u^*$ was found such that $\max_{0 \le u \le 1} J_B(u) = J(u^*)$, which minimized tumor mass and therapy administration, and maximized effector cell and IL-2 concentration.  A down-side to the model was that at the end of treatment, tumor mass tended to start regrowth.  Ghafferi and Naserifar \cite{Ghaffari:2010kx} improved upon the model of Burden at al. \cite{Burden:2004} by including a linear penalty, $-\omega y(t_f)$, where $\omega$ is constant, $y$ is the quantity of cancer cells and $t_f$ is the final time-point of observation, such that their objective functional is $J_G(u) = -\omega y(t_f) + J_B(u)$ (note that $-\omega y(t_f)$ is the terminal payoff for $J_G(u)$, as defined above).  The updated objective functional allowed for identification of a treatment that stops tumor regrowth and also acts faster than the treatment identified with $J_B(u)$.

Similarly, Castiglione and Piccoli \cite{Castiglione:2006uq} used optimal control for a model they developed to investigate dendritic cell vaccine (DCV) therapy for solid avascular tumors.  For $N$ total number of vaccinations and $\eta$ total time for one vaccination, the space of admissible controls, $U$, was taken to be 
\[U = \{u_s: S \in \mathit{S}\},\]
where $\mathit{S} = \{t_i: i=0,...,N-1, 0 \le t_0 \le T-\eta\}$  is the space of DCV injection schedules, $\mathit{S}$, and for every $S$ in the space of schedules $U_S$, $u$ was taken to be
\begin{equation}
u_s(T) = \sum_{i=0}^{N-1} \bar{u}(t-t_i)\chi_{[t_i, t_{i+\eta}]}, \label{eqn:hybrid}
\end{equation}
where $\bar{u}:[0,\eta] \mapsto [0,\bar{V}]$, with $\eta$ the total time of vaccination, $\bar{V}$ maximum vaccine amount, and $\chi$ the indicator function.  For every $u_s$, the therapy consisted of vaccine injections, modeled by $\bar{u}$, at time-intervals $[t_i - \eta, t_i]$ for all $t_i$ in $S$.   The objective functional (here termed the cost functional since it was minimized) was taken to be the final value of the tumor mass $M$, and the optimal control problem was stated as follows: for initial condition $x^0$, what is the schedule $S \in U_S$ such that $M$ attains a minimum?  The problem stated as such is known as the \emph{Mayer} form, which can be used when the running payoff, $r$, is zero in Equation \ref{eqn:Bolza}.  The authors took $T = $6 months, $N =10$ DCV injections, and $\eta = 0$.  An initial schedule $S_0$ was chosen randomly and 2000 optimization steps were taken to find the optimal $\bar{S}$, which resulted in almost complete clearance of a tumor.  Nevertheless, resurgence of tumor cells occured between vaccinations and the final value of the tumor mass was highly dependent on the timing of the last vaccination.

In order to correct for this effect, Piccoli and Castiglione \cite{Piccoli:2006} expanded the cost functional in \cite{Castiglione:2006uq} to also minimize the time during which the tumor is above a maximum mass, $M^{max}$, necessitating a switch back to the Bolza form for their objective functional.  The authors emphasized that the optimal injection schedule $u^*$ obtained is highly dependent on the parameter values and initial conditions of the system, indicating that the parametrization of the system from patient characteristics or other relevant knowledge should be performed carefully.  Castiglione and Piccoli \cite{Castiglione:2007ys} further expanded the cost functional in \cite{Piccoli:2006} to include, in addition to the previous constraints, drug holidays and minimization of vaccine injected to lower toxicity.  The authors also considered different types of cost functions: continuous, impulsive (similar to Equation \ref{eqn:hybrid}, but with $\eta = 0$), or hybrid (Equation \ref{eqn:hybrid}, $\eta \ne 0$), to take into account the different scales between duration of drug injection and tumor growth.  The hybrid method was found to be most effective for their system and the optimal treatment schedule that was identified consisted of a high initial dose and repeated smaller follow-up doses.  

These examples of optimal control theory applied to cancer immunotherapy show that the controls are developed slowly, with complex control functions often built upon more simple and established ones.
\subsection{Genetic algorithms}
For computational agent-based models, function optimizers such as genetic algorithms (GAs) are more suitable.  GAs are part of a class of evolutionary algorithms that work as follows: a set of $n$ initial binary strings, $\{S^{(1)}_i\}_{i=1}^{n}$, sometimes termed `chromosomes', representing possible solutions to the problem of interest are processed  via an evaluation function, $E_i = E_i(S_i)$, which gives a measure of the string's performance, and a fitness function, $F_i = F_i(\{E_i\})$, that compares the performance of all the strings to each other.  In the intermediate selection phase, strings $S^{(1)}_i$ are kept and duplicated with probability proportional to their respective fitness functions, $F_i$.  The strings of this intermediate population, $\{S^{(1*)}_i\}$ are then recombined with each other by cross-over and mutated at a low rate to obtain a new generation of strings, $\{S^{(2)}_i\}$ (note that the $i$ do not represent the same string between the generations (1), (1*), and (2)).  This process is repeated until an `optimal` string is found, either via running the algorithm for a fixed number of generations, or until a string is found that surpasses a threshold set by the fitness function.  This optimal string, like $\*u^*$ in optimal control theory, represents an optimal therapy solution given constraints imposed by the evaluation and/or fitness functions.  Importantly, an agent-based or other computational model can be run with any possible therapy ('string'), and the results can be used to determine the output of the evaluation function, thus the optimal string will give the optimal therapy vis-a-vis the computational model.  We have described the `canonical genetic algorithm' \cite{Whitley:1994} originally developed in \cite{Holland:1975}, and there exist many modifications that will not be presented here \cite{Whitley:1994, Mitchell:1998}.

Lollini et al.\ \cite{Lollini:2006vn} set out to use a GA to develop an optimal vaccine schedule for the agent-based SimTriplex model \cite{Pappalardo:2005kx} described in Section \ref{sec:class}.  The Triplex vaccine is a cell-based vaccine that stimulates an immune response via engineered HER-2/neu positive cells that express allogenic MHC Class-1 molecules (for increased recognition by CD8$^+$ T cells) and IL-12, which boosts both humoral and adaptive immunity \cite{De-Giovanni:2004zr}.  The GA was initialized with 80 binary 1200-bit strings, where each bit represents a time-step where a vaccine injection is/is not (1/0) given.  For each string, the SimTriplex simulator was used to determine mouse survival time ($s$) and maximum number of cancer cells in the transient ($N_{cc}^1$) and steady ($N_{cc}^2$) tumor growth phases.  The evaluation function was then taken to be,
\begin{equation}
f(n,s,\beta) = \frac{n^2}{s}\cdot \beta, \label{eqn:f}
\end{equation}
where $n$ is the number of injections specified by the test string, and $\beta = \beta(N_{cc}^1, N_{cc}^2)$ is proportional to the maximum number of cancer cells.  Note that in this implementation, the aim was to minimize the fitness function.  Via tournament selection (the fitness functions of two or more strings are compared and the probability of selecting a string is proportional to the fitness ranking of the string), followed by mutation and elitism (the best strings are not mutated), an optimal string (i.e. vaccination schedule) was selected.  Also, in this case, Equation \ref{eqn:f} was referred to as the `fitness' function, which is not to be confused with our definition above.

An extension of this fitness function was applied in the algorithm in \cite{Lollini:2006vn} which incorporated several instances of \emph{in silico} mice with varying immune backgrounds to choose the optimal vaccination schedule, $S^*$.  This schedule was applied to two sets of 100 virtual HER-2/neu positive mice and resulted in $~90\%$ survival rate (this is in comparison to a chronic protocol, which requires 4 vaccinations in a 2-week 'on' and 2-week 'off' protocol, over the course of at least one year, and has 100$\%$ success rate, but is not realistic for clinical development).  A previous 'trial and error' schedule was found to reduce the number of injections by $\sim 27\%$ over the chronic schedule, whereas the one identified by the GA reduced it by $\sim 42\%$, yielded excellent survival, and generated similar immune dynamics to the current protocol.  In a follow-up study, Palladini et al.\ \cite{Palladini:2010ys} showed excellent agreement between the \emph{in silico} predictions and \emph{in vivo} experiments, and made additional observations, such as the importance of vaccination density in the early phase of the immune response and dependence of vaccine success on age of the mouse.  GAs and evolutionary algorithms have also been used to identify potential epitopes for vaccine development \cite{Brusic:1998qf}, to develop an initial guess for a discrete optimal control problem based on the model in \cite{Castiglione:2007ys}, \cite{Minelli:2007}, and to solve a multi-objective optimization problem for a model of combination chemo- and immunotherapy \cite{Kiran:2013}.

By example of optimal control theory and genetic algorithms, we have shown how mathematical methods in optimization used in conjunction with appropriate models yield treatment schedules derived in a systematic, rather than experimental or computational `trial and error' method, that can be used not only to create optimal treatment schedules, but to also better understand immune response to treatment, such as the observations obtained from both methods above to the relative importance of early vs. late immunization protocols.
\section{Challenge: Design and identification of combination treatment regimes} 
\label{sec:comb}
Just as combination therapy using non-immunotherapeutic approaches is a mainstay in cancer treatment, combination immunotherapy either with just immunotherapeutic agents or with immune- and non-immunotherapeutic agents can and should be designed rationally to maximize treatment response \cite{Mellman:2011kx, Overacre:2015}.  Modeling can contribute to this process as it can aid in the development of a mechanistic understanding for effectiveness of combination vs. mono-therapies and make the search for an optimal combination therapy more efficient.  For example, de Pillis et al.\ developed a system of six ODEs for combination chemo- and immunotherapy that included tumor, NK, CD8$^+$, and white blood cells, as well as bloodstream concentrations of chemotherapy (doxorubicin) and immunotherapy drugs, the latter of which include tumor-infiltrating-lymphocyte (TIL) therapy and IL-2 stimulation \cite{dePillis:2009}.  The model extended an earlier model by de Pillis et al.\ \cite{dePillis:2006ly} to include more recent biological findings.  The effect of chemotherapy, $M$, was modeled using a saturation term, $1-e^{-\delta M}$, where $\delta$ represents the efficacy of $M$,  and the boost to CD8$^+$ activity by addition of IL-2 was modeled using a Michaelis-Menten interaction term, similar to what we see in Equation \ref{eqn:PK_E} from  \cite{Kirschner:1998fk}.  Cell-cell interaction terms were modeled using previously fit equations, often of Michaelis-Menten type.  For example, the CD8$^+$ T cell stimulation by the tumor was taken to be
\begin{equation}
\frac{dL}{dT} \propto  j \frac{T}{k+T}L, \label{eqn:L_rec}
\end{equation}
where $T$ and $L$ represent the tumor and CD8$^+$populations, respectively, and $j$ and $k$ are parameters.  Following \cite{dePillis:2006ly}, several parameters of the model, which collectively determine CD8$^+$ T cell efficacy at tumor cell killing, were fitted to patient-specific data.  The authors found that the success of combination vs. mono-therapy differed based on initial patient characteristics.  or example a patient with higher CD8$^+$ T cell efficacy would benefit more strongly from immuno- or combination therapy than a patient with low efficacy, since an injection of IL-2 for the latter group would result in more CD8$^+$ T cells, but not enough to increase overall anti-tumor response.  Notably, had the authors modeled an immunotherapy that altered these variables directly (such as injection of CD8$^+$ T cell costimulator agonists that boost CD8$^+$ effector activity), they would have obtained different success rates for the therapy combinations.  Nevertheless, the model can serve as a forerunner to models parametrized with patient-specific parameters that can then be used to identify optimal therapy combinations.

Targeted therapies, such as antibodies against the breast cancer cell receptor HER2 and EGFR, elicit a strong, adaptive immune response, raising the implication that combination immunotherapy with targeted therapy can achieve a synergistic anti-tumor immune response \cite{Vanneman:2012uq, Xu:2016kx}.  The humanized antibody to VEGF, Avastin (bevacizumab), is FDA-approved for a number of different cancers, and in addition to blocking the pro-angiogenic activity of tumor-produced VEGF, also blocks the immunosuppressive activity of VEGF, which includes suppression of DC maturation \cite{Johnson:2007zr}.  Soto-Ortiz et al.\ \cite{Soto-Ortiz:2016bh} hypothesized that a combination therapy of an anti-VEGF antibody followed by administration of DC cells would result in a strong anti-tumor response by the immune system.
The authors built upon a model that focuses on tumor-immune interactions \cite{RobertsonTessi:2012fk} to develop a system of 18 ODEs that include tumor, immune, and vascular endothelial cells, and a number of cytokines and growth factors including IL-2, TGF$\beta$, and VEGF (TGF$\beta$ and VEGF are both considered to be immunosuppressive).  The authors simulated treatment with injection of anti-VEGF antibody and/or unstimulated DC cells and analyzed the results numerically.  The simulations showed that for tumors with low immunosuppression and high antigenecity, which is represented by a parameter that models the strength of maturation of DCs after encounter with a tumor antigen, the immune system can keep the tumor at a small size and administration of either therapy can kill the entire tumor.  When tumor antigenecity was lowered and immunosuppression increased, the simulated tumor grew and vascularized without therapy.  The authors observed that for tumors with high immunosuppression, DC therapy alone was not sufficient to kill the tumor without additional modification of the immunosuppressive microenvironment, which is an example of a prediction for effectiveness of a mono-therapy vs. combination when given initial system parameters.  The authors made a number of other observations regarding the interdependencies of tumor growth, antigenecity, and immunosuppression that would not have been possible in a simpler model.  A prediction arising from this analysis was that there exists a therapeutic window for optimal effectiveness of different immunotherapies that is dependent on the tumor size and growth rate, and thus knowledge of these tumor characteristics can be used to design the optimal mono- or combination therapy for a given tumor/patient.  Moreover, synergy in combination therapy was predicted to occur when immunotherapy followed anti-VEGF treatment, thus the model was able to address questions of efficacy prediction and optimal time and selection of treatment, and hence effectively address all the major challenges we have discussed. \vspace{0.07in} \\
\indent Models for various combinations of traditional and immunotherapies have been developed \cite{Cappuccio:2007fk, dOnofrio:2012, Wilson:2012dq, Serre:2016zr}, which generally build on earlier mono-therapy, tumor-immune, or even combination therapy models.  For example, Chareyron and Alamir \cite{Chareyron:2009} extend earlier models developed by de Pillis et al.\ \cite{dePillis:2003b, dePillis:2005zr} to include chemo- and immunotherapy and use control theory to determine optimal therapy protocols, again addressing multiple big challenges discussed herein.   These models can give insights as to mechanism and efficacy of combination therapies in order to guide clinical development.
\section{Recommendations}
\label{sec:recs}
\subsection{Intracellular and multi-scale modeling}
Most current immunotherapy models are developed at the resolution level of cell populations.  However, modeling intracellular signaling cascades and metabolic responses in specific cell types can give insight into how therapies can act at the intracellular level, and what the relative contribution of different therapies is to cell response: intracellular vs. cell-cell.  As in the case of population-level models of immunotherapy, where the therapy component is often added to an existing, validated model of tumor-immune interactions, the intracellular models of therapy can build on known signaling pathways that interact with proposed therapies in tumors and T cells.  Intracellular signaling models have been developed for cancer signaling pathways for a variety of tumor types\cite{Hornberg:2006kx, Bachmann:2012vn}, and can be modified to include subpathways of interest, such as induction of PD-L1 or secreted immunosuppressive factors.  Saez-Rodriguez et al.\cite{Saez-Rodriguez:2007ys} built a Boolean network model of a signaling network activated upon T cell receptor and coreceptor activation and validated it using both literature and experimental wild-type and knock-out data that matched \emph{in silico} predictions.  This network could be made more specific for different T cells (CD4+, CD8+, Treg), receptors of therapeutic interest (e.g. CTLA-4, PD-1) and their downstream signaling targets, which can be added to the model in order to examine how therapy would affect T cell activation status.  Modification of networks identified in the literature may require additional experimental input, depending on the current availability of data.  Such models would allow a more granular systems-level analysis of mechanisms of pharmaceutical action and subsequent therapy optimization and combination.  Furthermore, since immunotherapy drugs can act on multiple cell targets, for example CTLA-4 is expressed not only on CD8+ T cells but also Tregs, development of multi-scale models that incorporate both the interactions of different immune and tumor cell types and their respective intracellular dynamics would have the power to elucidate the multi-scale mechanisms of therapy action.  Indeed, multi-scale modeling has already been identified as a important method for generating systems-level predictions of cancer progression and therapy effectiveness \cite{Deisboeck:2011uq, Wang:2015fk}.
\subsection{Addressing toxicity}
Immune-mediated toxicities, including retinal dysfunction, liver toxicity, and pancreatitis can occur upon administration of immunotherapy.  Toxicity, which is generally caused by targeting of self-antigens by the drug-strengthened immune response, is often associated with clinical response \cite{Amos:2011fk}.  Incorporation of immunotherapy-related toxicity into models can result in estimates of therapy dose that take into account toxicity-related effects, as was already done in \cite{Burden:2004}.  Development of more mechanistic models can help to optimize therapies to maximize effectiveness and minimize toxicity if mechanisms for each are not entirely overlapping.  This may be especially important in models of combination immunotherapy, where toxicity-related events tend to be more common than with monotherapies \cite{Melero:2015uq}.  One route of interest towards this goal can be a joint experimental-modeling focus on Fc-FcR interactions.  Immunomodulatory mAbs such as ipilimumab, a mAb to CTLA-4, are most often of IgG isotope: in addition to containing two antigen-binding (Fab) domains, they contain a fragment crystallizable (Fc) 'tail' that can bind Fc receptors (FcRs) found on effector cells such as macrophages and DCs.  Fc-FcR interactions have been shown to contribute to cytokine-related adverse events during mAb treatment, and hence modulation of these interactions during mAb design could help to decouple therapeutic from toxic responses.  A better mechanistic understanding of Fc-FcR engagement and response during treatment could help lead to rational design of mAbs that reduce adverse events related to Fc-FcR interactions (reviewed in \cite{Ryan:2016fk}).  By developing models of Fc-FcR action in concert with experimentalists, modelers can hasten this process and use their models to guide development of optimized therapy options.

\section{Conclusion}
\label{sec:conc}
Immunotherapy is one of the most exciting recent developments in cancer treatment, with new drugs coming on the market at an increasing rate, for an ever-larger number of cancers. But, as this review highlights, many unanswered questions remain, and the field faces several challenges.  Mathematical modelers have addressed these challenges as early as the 1980's (e.g., \cite{DeBoer:1985}). The models we have reviewed here and their use in understanding various aspects of immunotherapy crucial for effective treatment, show that mathematical and computational models can play the role of a key enabling technology to improve the application of this type of treatment. However, as this review also discusses, there is much work to be done.  Almost all immunotherapy-specific modeling efforts we could identify focus on the population level, even though it is clear that a multi-scale approach is needed that also incorporates the molecular scale.  To aid this process and to stay at the forefront of immunotherapy technology development, high-throughput data derived from technologies such as microarrays and RNA-Seq for cell-level transcriptional information and mass cytometry (CyTOF) for population-level marker distribution can be incorporated into model development \cite{Charoentong:2012fk}.  

Collaborations between modelers, basic scientists, and clinicians are crucial for the realization of the translational potential of mathematical modeling in the service of a precision medicine approach to this revolutionary new cancer treatment. As many of the models show, parameter values matter greatly in predicting treatment outcomes and devising optimal treatment approaches. Personalized models that can be calibrated with parameters characteristic of an individual patient will turn mathematical models into powerful tools that can play an essential role at the bedside, not just in the laboratory. 
\small
\section*{Acknowledgments}
AK gratefully acknowledges support from the National Cancer Institute of the National Institutes of Health postdoctoral fellowship award F32CA214030.  RL gratefully acknowledges support from the National Cancer Institute of the National Institutes of Health under grant 1R01CA188025-01. 

\bibliographystyle{plain}
\bibliography{JRSI_Refs}

\end{multicols}
\end{document}